Computer Supported Cooperative Work (CSCW)
https://doi.org/10.1007/s10606-018-9333-1

# The Types, Roles, and Practices of Documentation in Data Analytics Open Source Software Libraries

## A Collaborative Ethnography of Documentation Work

R. Stuart Geiger[1] , Nelle Varoquaux[1,2] ,
Charlotte Mazel-Cabasse[1] & Chris Holdgraf[1,3]
[1]*Berkeley Institute for Data Science, University of California, Berkeley, 190 Doe Library, Berkeley, CA, 94730, USA (E-mail: stuart@stuartgeiger.com);* [2]*Department of Statistics, Berkeley Institute for Data Science, University of California, Berkeley, Berkeley, CA, USA;* [3]*Berkeley Institute for Data Science, Helen Wills Neuroscience Institute, University of California, Berkeley, Berkeley, CA, USA*

**Abstract.** Computational research and data analytics increasingly relies on complex ecosystems of open source software (OSS) "libraries" – curated collections of reusable code that programmers import to perform a specific task. Software documentation for these libraries is crucial in helping programmers/analysts know what libraries are available and how to use them. Yet documentation for open source software libraries is widely considered low-quality. This article is a collaboration between CSCW researchers and contributors to data analytics OSS libraries, based on ethnographic fieldwork and qualitative interviews. We examine several issues around the formats, practices, and challenges around documentation in these largely volunteer-based projects. There are many different kinds and formats of documentation that exist around such libraries, which play a variety of educational, promotional, and organizational roles. The work behind documentation is similarly multifaceted, including writing, reviewing, maintaining, and organizing documentation. Different aspects of documentation work require contributors to have different sets of skills and overcome various social and technical barriers. Finally, most of our interviewees do not report high levels of intrinsic enjoyment for doing documentation work (compared to writing code). Their motivation is affected by personal and project-specific factors, such as the perceived level of credit for doing documentation work versus more 'technical' tasks like adding new features or fixing bugs. In studying documentation work for data analytics OSS libraries, we gain a new window into the changing practices of data-intensive research, as well as help practitioners better understand how to support this often invisible and infrastructural work in their projects.

## 1. Introduction

The work of collecting, processing, analyzing, and visualizing data increasingly involves programming, particularly when working with datasets at large scales and levels of complexity. This form of programming is quite different than modern software engineering, where teams of developers produce software applications for others to use. Contemporary academic researchers and data scientists increasingly work with data by writing scripts (Langtangen et al. 2006; VanderPlas 2016). These scripts are relatively short segments of code that load data, process it, and output a result or an intermediate dataset for further processing. Open source scripting languages like Python and R have gained massive popularity in data analytics in recent years, competing with commercial data analysis applications that provide graphical user interfaces (GUIs) around proprietary scripting functionality, like Excel, SPSS, SAS, or MATLAB. Scripted analyzes are also gaining popularity due to the open science and reproducibility movements, as the software is free to use, open to review and modification, and lets researchers share every step taken in processing and analyzing the data (Sandve et al. 2013; Wilson et al. 2014; Kitzes et al. 2018).

The rise of the Python and R scripting languages in academic and industry data analytics (increasingly called "data science") is also often attributed to the massive amount of specialized open source software "libraries" specifically for data analytics (Robinson 2017). Software libraries – sometimes called "packages" or "modules" – are curated collections of reusable code that programmers can import to help script a specific task. For example, loading a dataset from a file, computing a linear regression, and visualizing the regression on a scatterplot can all be done in less than 10 lines of Python or R, by relying on functions and objects imported from libraries like pandas, ggplot2, or Matplotlib. Without importing such libraries, these common tasks would require writing several orders of magnitude more code – and far more programming expertise. Writing scripts that rely on open source software (OSS) libraries lets analysts perform the same data analytics tasks as in traditional data analysis GUIs, but the actual work of 'using' a software library is closer to software engineering than the point-and-click model of data analysis GUIs.

Many researchers and analysts who do this kind of data work are not primarily trained as programmers, but are increasingly expected to program with these libraries as part of their data analysis work. Documentation is thus crucial for a library's usability, as it is the primary interface available to interact with a data analytics OSS library. In addition, the decentralized, peer-production nature of OSS analytics libraries means there are many competing and complimentary tools that an analyst could use to perform the same task, and each library contains dozens or even hundreds of functions. The analyst needs to identify, understand and learn each of the relevant tools to perform the task of interest. Finally, libraries are developed and maintained by many partially-overlapping groups in the broad open source community, continuously evolving based on individuals' contributions. Documentation is a crucial way for analysts to understand the functions of the multitudes of libraries available to them, as well as how to actually utilize the library to get a task done in their scripts.

While documentation plays a crucial role across the open source ecosystem, it is notoriously considered low-quality, sparely written, out of date, or simply non-existent — both in and out of the data analytics context. In a 2017 GitHub survey of OSS contributors, 93% reported that "incomplete or outdated documentation is a pervasive problem" but "60% of contributors say they rarely or never contribute to documentation" (Zlotnick et al. 2017). One survey of OSS contributors at a scientific computing conference found that "on average respondents believe they should spend roughly 20% more time on documentation" than they currently do (Holdgraf and Varoquaux 2017). The same survey also found that contributors neither enjoyed nor felt they received as much credit for writing and reviewing documentation as they did for writing and reviewing code. While our research focused on documentation for data analytics libraries, software documentation more broadly is an interesting and challenging issue for many reasons of interest across CSCW, the social sciences, computer science, and to practitioners. Documentation for OSS libraries is critical yet often-overlooked infrastructure behind the already critical yet often-overlooked infrastructure of OSS libraries more broadly.

In this article, we examine the changing practices of data analytics through the window of OSS library documentation, particularly focusing on the invisible and infrastructural work that takes place to produce and maintain these crucial resources. Focusing on the specific context of OSS data analytics libraries, we ask:

- What is "documentation"? What are the different formats, types, or genres of documentation?
- What roles do documentation play in these communities?
- What are the practices around producing and maintaining documentation? In particular:
    - What skills are involved in doing documentation work?
    - What barriers exist to doing documentation work?
    - What motivates contributors to do documentation work?

We find that documentation for OSS data analytics libraries are in a variety of formats and genres, which reflects a wide set of roles that documentation plays—including but going far beyond the pedagogical role of informing data analysts about how to import and use particular functions to do a task. Documentation also serves an important public-facing role for analysts who are deciding whether to use a particular library, as well as serving key organizational roles for the developers who maintain such libraries. The multiple overlapping forms and roles of documentation mean that the work of producing and maintaining documentation is similarly complex. In this context, documentation work involves a wide range of skills, and documentation contributors face a variety of technical and social barriers. Finally, the motivations of those who do documentation work are similarly multivalent. We find that contributors' motivations range from a few who report as much inherent enjoyment in documentation work as they feel in writing code, to many more who describe documentation as a difficult or tedious chore they do for the good of the library.

1.1. Scope and method

*1.1.1. Research context*

We have two audiences for this paper: 1) CSCW researchers who find open source software documentation to be a rich phenomenon for investigating classic issues in the field and 2) practitioners in OSS communities who are interested in tackling the issues around writing documentation. We focus on the context of open source data analytics libraries, though we believe that the issues in this community have implications for the broader open source community as well. As in Ribes and Finholt's (2007) discussion of long-term research infrastructures, to serve both these audiences, we identify a broad range of issues and tensions that apply across cases. As OSS communities can be quite different—with a variety of norms, structures, goals, resources, scopes, and challenges—solutions to problems around writing documentation need to be tailored to local issues, circumstances, priorities, and needs. To support this, we identify thematic areas for further investigation, exploration, reflection, and intervention by both researchers of documentation and those who write OSS library documentation as part of their life and work.

Our dual audiences reflect our authorship, as this research is a collaboration between two ethnographers embedded in an institute organized around data science and two academic researchers at this institute who are also active contributors to data analytics OSS libraries. The primary empirical material for our findings in this paper are a set of semi-structured interviews with open source software contributors. This project is also empirically grounded in the ethnographers' broader fieldwork and the researchers' extensive experiences and work in this space. The ethnographers have been engaged in two years of fieldwork—including participant-observation, interviews, and trace ethnographic research (Geiger and Ribes 2011)—around various activities and topics in the computationally-supported production of knowledge.

This project specifically emerged out of a hackathon-style event around documentation organized by two of the co-authors of this paper called the "Docathon," which the two ethnographers were studying as a time-bounded collaboration event. As the research unfolded, the four of us realized our common concerns were around a much wider variety of issues about software documentation in OSS communities. We shifted analytical and methodological frames and moved into a more collaborative project. The ethnographers contributed expertise in qualitative interviews and inductive grounded theory analysis, and the scientists/developers contributed their lived experiences, perspectives, and sensitivities around writing documentation for (and participating in) OSS library communities. This project is therefore a part of a "collaborative ethnography" as defined by Lassiter (2005)—not just a collaboration between ethnographers, and not just ethnographers giving back to the people they study, but a "process [which] yields texts that are co-conceived or co-written with local communities of collaborators and consider multiple audiences outside the confines of academic discourse, including local constituencies."

*1.1.2. Methodological details*

The Docathon was organized at a research institute at a large research university in the U.S., which was founded in part to support the development of open and reproducible data science infrastructures and practices. The event was a hybrid co-located and virtual event, involving co-located participation at the main institute, satellite participation at two partner research institutes in the U.S., and remote participations via the Internet. Many participants were existing contributors to OSS data analytics libraries who came with specific documentation-related tasks, including: writing new documentation for undocumented features; writing new genres of documentation (e.g. introductory tutorials, galleries of examples); improving, updating, or reorganizing existing documentation; and developing software to make documentation work easier. Other participants were new contributors to these OSS projects, seeking to get involved in these projects by helping with documentation work. All participants interacted through mediated channels, including a mailing list, Slack chat channels, and GitHub repositories. The ethnographers' empirical work included: participant-observation fieldwork at the Docathon; interviews with Docathon organizers and participants (both at the main co-located site and those at remote sites) at the end or after the event; and interviews with contributors to open source software libraries who did not attend the Docathon. All interviews were semi-structured and based on a set of topics, themes, and issues that the four of us collaboratively generated, based on combining our different expertises and perspectives on documentation. Interviews ranged from 30–90 minutes, with most between 45 and 60 minutes.

We inductively analyzed interview transcripts for themes in a multi-stage grounded theory approach (Glasser and Strauss 1967), creating three hierarchical levels of thematic codes. We identified common themes around specific re-occurring expectations, frustrations, concerns, goals, barriers, as well as coded for specific data analytics OSS libraries mentioned to compare interviewees' experiences within and between projects. In our coding, we found that 'tensions' was a useful way of conceptualizing the different trade-offs, and re-coded interviews with pairs of thematic codes such as "simple versus complete" or "newcomers versus experts". We conducted additional interviews to further explore themes that emerged as particularly relevant for issues in documentation, like motivations, as well as to get perspectives from individuals who were not in our original round of interviews. We also shared our preliminary findings with various OSS developers and users to receive feedback and further iterate on our analysis and discussion, including facilitating a 1 hour discussion about documentation at a data science conference.

The 11 interviewees included the two Docathon organizers, 5 co-located Docathon participants, 2 remote Docathon participants, and 2 non-participant OSS contributors. Interviewees generally had substantial experience writing code and contributing to open source software in general, but a smaller number were newer to open source. Most had some experience writing documentation for OSS libraries, and as interviews were conducted at the end or after the Docathon, all had contributed to OSS documentation in some way. All of our interviewees are either current or former academic researchers who currently use open source software as part of research, either working in academic or industry settings. They were overwhelmingly but not exclusively male and either North American or European, reflecting levels of participation in open source in general. Our interviewees had strong variance in national origin and native language, with several non-native English speakers.

### 1.2. Literature review

#### 1.2.1. *Documents as social and organizational practice*

While the role of documentation specifically in OSS libraries is a less- studied topic, documentation in organizations and research has long been studied across the social sciences, particularly in CSCW, organization studies, Science & Technology Studies, and library and information science. The multifaceted work around documentation has been a longstanding concern in librarianship (Briet 1951; Buckland 1997) and in ethnographic studies of science and engineering labs (Latour and Woolgar 1979). In studies of workplaces and formal organizations, researchers have discussed how people use various genres of documents — including records, forms, interoffice memos, e-mails – to "accomplish and co-ordinate their day-to-day practical activities" (Luff et al. 2000, p. 12). For newcomers, learning how to properly read and write documents in a particular organization is a core part of learning how the organization operates, how different parts of the organization relate to each other, how decisions are made, and by whom (Darville 1995; Geiger 2017). As Trace (2011) reviews, scholars have used frameworks such as distributed cognition, activity theory, actor-network theory, coordination theory (Crowston 1997), and practice theory (Osterlund and Paul Carlile 2005) to discuss the role of documents in both collaboratively getting work done and sustaining the structure of an organization.

Ethnomethodology (Garfinkel 1967) and symbolic interactionist approaches (Goffman 1959) each emphasize how people produce and maintain social structures through descriptions, both spoken and written. Documents are an important way in which people make their understandings and intentions known to others.

As descriptions and documents are concrete, material externalizations of more complex or situated phenomena, they do not just describe the phenomena but also "enact" it. Documentation work — which includes reading, writing, editing, reviewing, organizing, circulating, searching, archiving, and destroying documents—is often a mode of collective sensemaking about not just the specific task at hand, but how the task fits into the organization's broader goals and principles. In a classic infrastructural paradox (Star 1999), documentation work is simultaneously a crucial part of an organization's operation and an often-overlooked form of invisible work. Extensive work has been done in hospital settings, for example, discussing how relationships between patients, nurses, technicians, and doctors are differently mediated through different systems and practices for recording patient records (Berg and Bowker 1997; Bowker and Star 1999).

### 1.2.2. *Software documentation and usability: technical communication*

Software documentation is a different genre of document than most documents that circulate in organizations and are typically studied by researchers from these theoretical traditions. Software documentation is often discussed in terms of topics like usability or instruction. Researchers in fields like technical communication have long focused on how to best write documents that communicate a product's features and functionalities to users (Weiss 1985; Van der Meij 1995). We instead discuss the more social, organizational, and infrastructural roles that documentation plays in software projects, particularly for open source projects. As we found, documentation often plays key roles in a project's internal processes, and is one of the primary mechanisms by which the project interacts with those outside of the core team. As such, it is a site of rich inquiry into the practices, processes, and values of software development and open source communities.

### 1.2.3. *Open Source Software from a communications and social media perspective*

Researchers both in and out of CSCW have also extensively studied the ways in which open source software developers communicate with users, particularly looking at modes of communication that go beyond the official codebase and documentation. These include user-generated content platforms (Storey et al. 2017), including developer blogs (Parnin et al. 2013), Twitter (Singer et al. 2014), YouTube coding videos (Poche et al. 2017), mailing lists, and question and answer (Q&A) sites like StackOverflow (Zagalsky et al. 2016). Story et al. (2017) also provide an extensive review of scholarship on how software engineers have used various media channels to interact with users, from the mid-1960s to today. This line of research has generally found that there are many different entrypoints and modes of communication between developers and users, with a wide variety of genres, conventions, and dynamics of interaction. OSS developers must serve a wide variety of users who have different levels of expertise and expectations, which is a challenge for both developers and users. We find similar challenges in our research.

*1.2.4. Organizational roles of documentation within developer teams*

In CSCW and organizational studies, researchers have extensively studied the roles that documentation for software or other technologies plays in formal, professional organizations (Hovde 2000; Vaughan 1997; Sellen and Harper 2003; Cohn et al. 2009). Researchers have also looked specifically at how software developers use trace data and textual artifacts like ad-hoc code comments (Storey et al. 2008) or code repositories like GitHub (Dabbish et al. 2012) to make sense of their work and the organization or community developing the software.

A long line of research has focused on the role of software documentation in professional firms that develop software for clients, with work in CSCW specifically focusing on the role documentation plays internally in a firm's software development process. For example, Cohn et al. (2009) studied 'agile' software development firms, in which rapid iteration is favored over the kinds of detailed textual planning documents that are more common in classical 'waterfall' software engineering. They found that despite the heavy anti-documentation rhetoric common among agile proponents, such firms still make use of textual artifacts to coordinate and plan software development, but use them quite differently.

*1.2.5. Peer production communities*

In the data analytics OSS libraries we observed, all were produced in a community-based, peer-production model, rather than exclusively by a single software engineering firm. As such projects are relatively decentralized communities, our research is also in conversation with scholarship on other OSS projects that follow this model, as well as communities like Wikipedia. Our study found various issues around topics extensively discussed in these literatures, including incentives and motivations (Balestra et al. 2017), onboarding newcomers (Steinmacher et al. 2015), social interactions on collaborative platforms (de Souza et al. 2016), work distribution and decentralized management (Gousios et al. 2016), and the overall project health and success (Crownton et al. 2003; Bangerth and Heister 2013). Such research has extensively discussed tensions between centralization and decentralization in peer production communities. Such projects benefit from the flexibility that comes from the lack of a formal organizational structure (especially early on), but this decentralization often comes with its own costs and challenges. Peer production projects often face growing pains as they grow and scale (Halfaker et al. 2013), seeking to organize and standardize in ways that let them achieve specific goals, while maintaining the consensus-based model that is seen as a core value in such communities (Tkacz 2014).

## 2. Findings

### 2.1. The many faces of documentation

#### *2.1.1. Defining multiple forms of software documentation*

A major issue around documentation is that it has several definitions. In our interviews, "documentation" was used to refer to a broad set of textual resources, rather than a single kind of text. From those interviews, as well as observations during the Docathon, we identify several major types of documentation. These types are not mutually exclusive categories, but they often have different intended audiences, conventions for presentation, skills needed, and formats for distribution. We list each in Table 1, and discuss them in more detail below.

**User documentation**, also called **narrative documentation**, typically gives a broad, high-level overview of what the library is intended to do, how to install it, or how to use it (see Figure 1). It is typically not an exhaustive list of everything the library does, and often targets new users or those who are still deciding if they want to use the package. It may include material that is not kept in raw text files (e.g., Jupyter notebooks, repackaged presentations, or videos), but is generally officially created by a project's developers and hosted on the project's webpage/repository.

**Galleries and examples** generally lack high-level motivations and structure, and instead present a short and specific outcome generated by a single block of code (Figure 2). They are typically created officially by a project's developers and hosted on the project's webpage/repository. Because galleries and examples are self-contained code, it is possible to run this code when the documentation is built in order to generate output figures (using a framework like sphinx-gallery[1]).

**API documentation** (sometimes called "**docstrings**" in the Python community) is text included in code comments at the beginning of functions or methods (See Figure 3). API documentation has a specific structure that can be parsed by libraries (such as Sphinx or Doxygen) which render it into structured output, like HTML pages. It typically includes a brief, high-level description of what the function does, followed by more structured information about the parameters the function uses. Many development environments (such as Jupyter Notebooks or RStudio) can interactively render API documentation to users and developers.

**Non-traditional documentation** While the types of documentation mentioned above are most common and well-defined, there is a wealth of unofficial or unstructured material on the internet that several interviewees mentioned. This includes content distributed with the code itself, such as well-written error and warning messages. It also includes distributed content that isn't created by the project core contributors, such as blogs, community Q&A sites like StackOverflow, or Jupyter notebooks. As these sites often rank highly in search engine queries, they are an important venue for learning and instruction. However, as documentation, they remain ad-hoc, unorganized, and rarely under the editorial control of a project's developers.

---

[1] https://sphinx-gallery.readthedocs.io/en/latest/

*Table 1.* Different types of documentations and how they are used.

| Type | Usage | Format |
| --- | --- | --- |
| User/narrative documentation | Aimed at users. Includes high-level summaries of the library's features, installation guidelines, tutorials, textbooks, and "getting started" or "quick start" guides. | html & pdf |
| Galleries and examples | Aimed at users. Consists of short, self-contained scripts that detail how to perform actions using the software library. Can be tested through executing examples. | html, code |
| API Documentation | Aimed at users. Provides detail the operation of functions along with available arguments and parameters. Can include code snippet examples. Can be unit tested. | html, pdf & interactive |
| Developer documentation | Aimed at contributors to the library. Includes guidelines for contributing and technical information about the project, such as automatic build systems for testing code. | html & pdf |
| Non-traditional documentation | Aimed at users. Includes error messages, social media platforms, Q&A sites, and other ad-hoc instructions about using the library. Many participants were unsure if these should be considered "documentation". | Various formats |
| Comments in source code | Aimed at contributors to the library. Ad-hoc explanations of how the library's code works. Was not mentioned as "documentation" by any of our participants. | Source code |

*2.1.2. Relationships and tensions in definitions of documentation*
In practice, there is an interplay between the above-mentioned types of documentation. One of the Docathon organizers described the difference between the three major kinds of documentation as ranging "from the most zoomed in to the most zoomed out" (Docathon organizer 2) – (from API documentation, to examples and

The Types, Roles, and Practices...

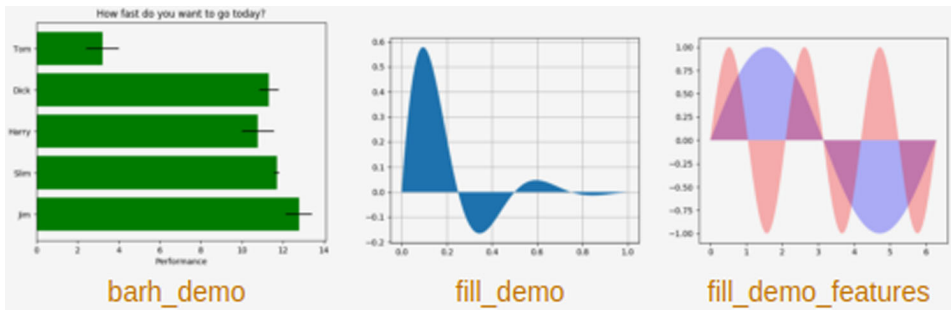

Figure 1. A form of user documentation: the scikit-learn "quick start" guide, which introduces the software through a narrative introduction to machine learning.

Figure 2. A section of the Matplotlib gallery. In Matplotlib, each plot links to separate HTML pages that detail how to reproduce the figure shown in the gallery.

Figure 3. API documentation for the value_counts pandas function, detailing how to call the function with specific parameters. Displayed in a webpage.

galleries, user/narrative documentation). One of the Docathon participants drew a comparison between a textbook versus a dictionary:

> [there is] a static version that basically says, 'Hey, this is a project. Here is what the project is meant to accomplish. Here is the project, then type this, the project starts, and then it can do this, this and this' [...] You basically can go through tutorials like you would read a book that tells you how to do statistics or how to do something else.

> The second kind of documentation [...] is basically, 'Here is a list, alphabetical order, or another order, of all the things the project can do. If you want to know how to use a function in particular, how a specific piece of code, you go to this subsection.' And this subsection will often be relatively short and tell you why and how it can be used and what it is related to, more like a dictionary. (Docathon participant 7)

These types of documentation can co-exist with one another, but they can also introduce tensions within the developer team and broader community. At the highest level, tension arises from an imprecise definition of what documentation means to a project. If someone is told to read "the documentation" or says that "the docs" need improvement, it can be unclear which of the above-mentioned types they mean. An additional tension arises when a project does not diversify the types of documentation they provide. As as each type of documentation has different goals, scopes, and audiences, conflicts can emerge if documentation is exclusively imagined as one of these types. Many interviewees noted that it was important for a software project to have good documentation across many different levels, bringing up examples where projects needed to work more on one specific type.

While we found that these kinds of documentation were often conceptually clear for interviewees, they were sometimes combined and merged in practice. One Docathon participant discusses the documentation for a smaller project they work on, where documentation takes the form of a single README document, with different subsections that do different kinds of work:

> [our software package] doesn't have a very thorough documentation, just a README, but it's a mix of everything. It's like, high level motivation, it has specific examples, and it has, how to install this thing, how to run it. It's kind of a very technical thing, so it targets more [...] hardcore developers. (Docathon participant 5)

Successful projects intentionally adopted a broad definition for what it meant to "contribute to documentation". For example, the Docathon's organizers opened the week with talks and tutorials that introduced the different types of documentation and discussed best practices for writing each type. They then encouraged participants to choose whatever definition they liked for the week. In our interviews, Docathon participants frequently made implicit and explicit use of these distinctions when talking about their work. Many chose to specialize in one particular type of documentation for the week, but each of the three major types of documentation had at least one person working on it.

Interviewees discussed moving between these formats. For example, one participant created a new static tutorial using slides they had created for an in-person bootcamp they taught. When asked to talk about examples of good documentation, many participants also praised tutorials that collectively worked as textbooks for a broader conceptual topic (such as machine learning). As one of the Docathon organizers stated:

> If you get enough of those tutorials together, then the documentation becomes some sort of [...] textbook [...]. It's like a collection of tutorials that will cover the space of ideas that this package cares about, which is, in my mind, something different from just having a collection of random tutorials because it starts to resemble something that's more similar to a traditional academic textbook or whatever. (Docathon organizer 2)

## 2.2. Roles of documentation

One reason for identifying the many types of documentation described above is that they are linked to the diverse roles that documentation plays in the community, as well as the end-user to which documentation is directed. Most software projects have many different kinds of documentation at once, each with its own relationship to the community.

Interviewees discussed how documentation helped with a variety of tasks, including: facilitating learning and education, giving a project publicity, serving as a signal of health, serving as external memory or a living document, facilitating testing and verification, onboarding newcomers to open source projects, and facilitating collaboration between developers. We found that interviewees highlighted subsets of these roles for documentation in their projects, though they were not usually cleanly tied to a single *type* of documentation. Because these roles are often hard to identify or define, tensions can emerge between community members who may have different expectations about what documentation is meant to do. The following sections describe common roles that documentation plays in the community.

### 2.2.1. Learning/education

The most agreed-upon role for documentation is as a pedagogical resource for people to learn how to use a piece of software. In this role, different types of documentation can be targeted at different audiences: an expert may use it to look up the details of a function, while a novice may need to look up whether such a function exists at all. The goal of learning was frequently contextualized and specified by interviewees, who discussed different kinds of learners and stages of the learning process. An often-imagined audience of documentation was someone searching for a piece of software to help them do a particular task:

> You can imagine a user, with some sort of need, Googling around trying to find some sort of software to do what they want to do. Then they happen upon software and try it. There's this patience period that probably is something like five minutes, during which they may try a software. Then it might not work, probably won't work. Then if there's no documentation to help, that user is basically lost for that software project and will say, "I tried that but it didn't work." You need documentation, like ideally of everything but especially of the very beginning of, to create a minimal user experience and have it in the documentation how to set the thing up and how to do the thing that it's supposed to do. (Docathon participant 6)

Interviewees frequently discussed forms documentation like tutorials or galleries as intended for new or potential users, while API documentation and docstrings were for those who were already using a particular piece of software. One Docathon participant discussed these differences in answering a question about how they use documentation in their own day-to-day work:

> I use the docstrings all the time, a lot of this through interactive work [...] even for simple things like, what is the order of the arguments of this function? [...] Examples are pretty useful when I get started with things with the new software that I haven't used before. [...] I was looking around for software to model, do statistical modeling of longitudinal studies. I started looking at [...] a Python project, and I was actually bounced off of that because there were very few examples, none of which looked like what I was trying to do, so I couldn't get that. (Docathon participant 9)

### 2.2.2. *Publicity/signal of health*

The above quote shows the overlap between documentation as a resource for learning and a second role: as an advertisement for the software project. In much of the open source software ecosystem, there are overlapping and competing projects, and we frequently heard mention of documentation as a reason for deciding which project to choose. This was true both from end users (who discussed deciding about whether to use a piece of software based on its documentation) as well as project maintainers (who discussed improving documentation in order to recruit new users). One of the Docathon participants discussed making such decisions as a user, which was an unprompted response in a longer answer a question about how they use documentation:

> I love documentation, I use documentation all the time. In fact, it's certainly the case I decide whether to use a project or not based on the quality of the

> documentation [. . .] If I'm looking for a library that does something and I have, you know, five libraries, there are different criteria that I use to decide which one I'm going to use but quality of the documentation is certainly one of them [. . .] (Docathon participant 5)

One of the Docathon organizers discussed this issue with a software project they help maintain, where the team had previously worked to overhaul the project's documentation:

> that was the biggest scale project that I worked on in terms of documentation [. . .] it was clear by the end of it that when you looked at the website after that overhaul had happened [. . .] there was a clear value added to the project. Even though none of the code of the actual project itself had really changed, it was just, again, the presentation of the ideas surrounding that code base. It made it much easier for me to discover other parts of the package that I hadn't learned about already, and also made it much easier for me to pitch it to somebody else if I was like, "Hey, you should try [software] to do this stuff." When I could show them that website, it was clear that the project was well-constructed and well-managed and had its act together. (Docathon organizer 2)

### 2.2.3. Institutional memory/living document

Many of our interviewees are long-standing participants in open source software projects, including several who have spent years as core maintainers of projects with dozens or even hundreds of contributors. In such projects, documentation plays an important organizational role as a living memory for the project that records every change made. We often heard from interviewees about how projects are difficult to manage at such large scales without good documentation practices. For example, core developers mentioned having difficulty remembering what changes had been made after a dramatic refactoring of the code. Interviewees also spoke about several cases where an old feature was unused because there was no official documentation written about it, and the only way to discover its existence was to look through the code itself.

One Docathon participant (who is a core developer for several software projects) was asked why they write documentation, responding first by saying: "Well, one, because I forget how things work. That's the most valuable thing from my point of view." They then discussed the importance of "making the product usable." (Docathon participant 5). Another Docathon participant (also a core developer for several software projects), when asked about how much documentation they write, stated:

> I've been doing it more and more recently . . . I care more and more because I come across more and more things I've written a couple of years ago, and I have no clue what I fricking wrote (Docathon participant 7).

*2.2.4. Reference point for collaboration between developers*
Aside from serving as institutional memory, we found that documentation also facilitates collaboration between developers of a project. Many OSS data analytics libraries are modular collections of different functions that are developed relatively independently from each other (compared to more traditional software applications). As documentation summarizes the overall design of a feature, module, or function, some interviewees spoke about how good documentation can be a useful reference point for developers to communicate their ideas and intentions to one another. For example, one interviewee who maintains a large, complex project (both in terms of number of contributors and number of features) discussed how existing API/reference documentation is sometimes referred to in discussion threads about proposed new features or refactoring existing features. They noted that because many developers restrict their contributions to a small part of the library, discussions about large-scale changes to the code are facilitated by linking to the API/reference documentation. They noted that when attempting to describe where changes to features/APIs should be made, "if there's already relatively complete documentation, that's very easy to describe in a single email" (OSS contributor 10)

*2.2.5. Testing/verification*
Documentation was also described as an important part of coding itself, particularly in testing and verification. This takes several forms, starting with API documentation as a way for developers to externalize their intentions by describing what they want a function to do. Some interviewees had established practices of writing documentation as they wrote code. One compared this practice to unit testing, which is used to ensure that key functionality of the package had not changed. As one of the Docathon organizers explained:

> ... it's trying to give the user an intuition on what the method does. [...] it also allows me to make sure that I understand exactly when the method works and when it doesn't work. [...] it also allows us to check that the API is nice, and it's also a very simple way to check that the method works. So this is also very common in research, you just look that things make sense, and sometimes you don't really get this information when writing unit tests. (Docathon organizer 1).

*2.2.6. Onboarding newcomers to open source projects*
Finally, a major auxiliary role that documentation plays is as a way for newcomers to contribute to open source projects. In open source software communities (both in and out of the data analytics context), documentation has long been discussed as a kind of low-risk, entry-level task that will help newcomers gain familiarity with the project—a model long discussed by scholars of legitimate peripheral participation (Lave and Wenger 1991). In fact, the Docathon organizers reported that one of the key reasons for organizing the event was to connect open source software projects in need of documentation work with people who wanted to get involved, but were unsure how.

Documentation work is seen as a good task for onboarding, because newcomers can work through the process of submitting changes for review (e.g., a GitHub pull request) without having to also advocate for a change to the codebase. Another reason for newcomers to work on documentation is that they are often in the best position to know what is confusing, unclear, or important to someone new to the project. However, it should be noted that having a fresh perspective is often a trade off with being able to contribute high-quality documentation in line with the project's standards and goals. One of the Docathon participants, who also is a core contributor to many open source software projects, summarized some of the major benefits and drawbacks:

> Interviewer: One thing that some people have suggested is that documentation is a good place for people who are new to open source to get started. How do you feel about that?
>
> Docathon participant 7: I would agree and disagree. I would agree because it's relatively easy to start contributing to. You don't need to understand the code. It's really nice when you're new to open source, and you need to understand the process of submitting patches. You don't have this overhead of thinking about, "Is the code I'm writing correct? I can focus on the workflow." [...] It makes it great also because if you're new to a project you have the views of newcomer, and so you realize what is not of use from the documentation [...]
>
> The problem is, to write good documentation you need to already have, usually, I think, relatively good knowledge of the project, because you need to understand how pieces are intertwined. [...] And how they interact with each other and what are the useful and useless information or the thing that may be missing. Which, by definition, someone who is new to a project cannot know. At the same time, once you're familiar with the project, you don't see anymore what's needed for a newcomer. So it's both the right place and the wrong place to start in my opinion.

One Docathon participant had used open source tools, but had never contributed to an open source project. They to the came to Docathon specifically to start contributing, and reported a generally good experience:

> Docathon participant 4: I think that the Docathon was a great low-barrier way of getting acquainted with how it all works. [...] docs are something that anyone can sort of critique and improve, even if they don't necessarily have a deep knowledge about the code base.

However, we also found pain points and lessons learned in using documentation for onboarding newcomers. Some Docathon participants who were newcomers to a project were not able to easily know what tasks needed to be done, and did not want to make substantial changes to documentation without specific guidance. While changes to documentation are often easier to get approved than changes to code, interviewees recounted many conflicts over documentation, including those involving newcomers (see Section 2.3.3 on documentation standards). Several interviewees discussed having to go back and forth with what was presumed to be a non-controversial update to documentation, sometimes waiting days to get a change approved. In some cases this was because of a lack of input from the core developers, in others because of direct disagreements over how the contribution should proceed.

In all, we found that documentation can be a productive and low-barrier way for newcomers to contribute to open source software projects, but we emphasize the need for projects to actively support such forms of peripheral participation. While we leave a systematic study of onboarding for future research, we find more cases of success with projects that have a well-developed culture of documentation, where there are clear and agreed-upon standards for documentation, active review procedures, and where most core contributors to code also contribute to documentation (we also discuss tensions around who does documentation work later). Our interviewees mentioned several projects that have such qualities (with relatively consistent answers), as well as many more that did not.

2.3. Skills and barriers around documentation work

Developers and users of data analytics OSS libraries generally acknowledged that software documentation is important, yet documentation is routinely either not written or not kept up to date. Like all the issues and tensions discussed above, there is a spectrum: some projects have little to no documentation in any form while other projects highly prioritize documentation and integrate its development into community practices. However, documentation seems to be consistently under- maintained in such projects, by contributors' own standards. For example, a recent questionnaire asked contributors to open source scientific Python libraries to state what percent of their time they think *should* be spent on documentation, versus what percent of their time they *usually* spent on documentation (Holdgraf and Varoquaux 2017). There was a general distance between these two responses, reflecting a belief that open source developers felt they should spent more time on their project's documentation (See Figure 4). These findings are further supported and contextualized by our ethnographic and interview research, in which contributors routinely discussed a wide range of issues around why documentation work was both a personal and collective challenge. In the following section, we discuss skills involved in documentation work, technical barriers that often exist for contributors, and issues around standards and quality for documentation.

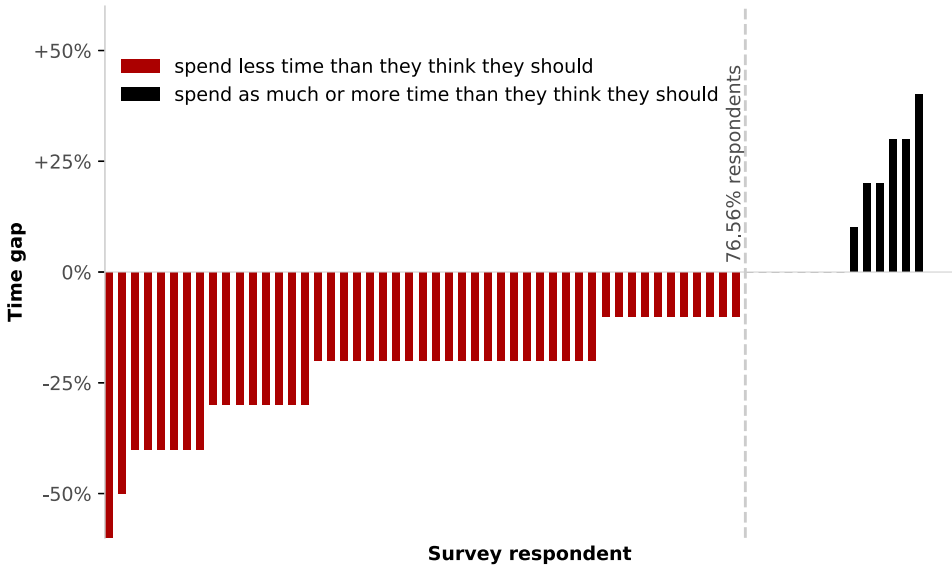

*Figure 4.* The gap between how much surveyed OSS library developers *usually* spend on documentation versus they time they think they *should* spend on documentation. Each vertical bar is a single respondent from a survey of developers at a scientific computing conference. Negative values represent respondents that believe they should spend *more* time on documentation, codified as "Documentation Guilt." Figure created from data by Holdgraf and Varoquaux (2017).

### 2.3.1. Skills

The most straightforward barrier to creating good documentation is having the skills to do so. Contributing code to open source software requires a specific set of skills: knowledge of the programming language used, version control, and other practices in software engineering. Writing, contributing, and reviewing documentation often requires not only these skills but also an additional set that are often not taught in traditional software engineering. These include communication skills, creative writing, empathy, and good knowledge of the English language (which for some contributors may not be their native language). Below we identify several skills important to documentation.

*Self-efficacy* Like all skills, individuals must have both have the skill itself and self-efficacy, the belief that they have such a competency. Throughout the interviews, many participants expressed that they lacked the correct skills to write good documentation for their own software.

> I don't know many people who enjoy writing documentation. I think one of the reasons being it's not a skill that we learn very well, so I think a lot of us feel that it's not something we're good at. If we have been feeling different, that we're good at it, probably we would enjoy it more, but it's sort of a painful process to do. (Docathon organizer 1)

*Empathy* Some interviewees expressed a tension around their status as advanced users and developers, as documentation is often seen as used primarily by novices. One concern expressed by some interviewees is a lack of empathy, as documentation work involves putting oneself into the user's shoes, and advanced developers may not know what their audience actually needs:

> you need to have a very good sense of who your audience is, and what you need to tell them when. [...] The biggest problem is that what I need in documentation is not necessarily what someone coming to the library using documentation does. I may be lacking sufficient empathy to write what newcomers need. Whereas a newcomer probably still remembers what they didn't know yesterday and can write the docs with that in mind. (Docathon participant 3)

*Language proficiency* Interviewees discussed the importance of various communication skills, which go far beyond the skills required to fix a bug or write a new feature. For example, many interviewees felt that far more English skills were required to write documentation than to write code or even informally interact with others in the project. Several of our respondents were not native English speakers, and many respondents said that they had observed this barrier in projects:

> It [writing documentation] actually requires writing much more English than code requires. They don't necessarily feel as competent to do that, or they ask for help. (Docathon participant 9)

*Communication skills* While being competent with a language is a pre-requisite in community interaction, it is not enough to guarantee strong documentation. It is also important to be able to communicate ideas in an easy-to-understand manner. Documentation is intended to be public material to be read by an audience, and many of our respondents emphasized how storytelling and creative writing skills were highly important for documentation:

> Creative writing is important, to enable search to boil down whatever are the key features of the software, and also what the science of the software is doing, down to clear explanations (Docathon participant 9)

*Knowledge of software to be documented* Finally, interviewees discussed how documentation contributors also need a good working knowledge of the software library being documented (and the concepts behind it) in order for the documentation to be accurate, precise, and concise. This does not refer to the technical barriers of participating in open source software communities (e.g. how to use GitHub), which we discuss in the next subsection. This can be in conflict with an increasingly popular trend in some open source software communities in which newcomers are encouraged to write documentation before contributing code. As we previously discussed, it is important to understand how the process of writing documentation is a collaborative effort between experts and newcomers.

### 2.3.2. Technical barriers

In addition to the skills involved in writing documentation discussed above, there are often substantial technical skills required to contribute this work to an open source software project. Projects frequently store documentation in the repository they use to store code, requiring a working knowledge of version control and online code repositories like GitHub. While contributing documentation is an increasingly popular onboarding mechanism, it often challenges new users with skills and workflows with which they are not familiar.

Furthermore, with many forms of API documentation (like docstrings), the documentation text is stored as comments in the code itself. This means that contributing documentation typically follows the same complex process and workflow as contributing code: downloading the code repository, installing it on one's own computer, adding or editing the documentation text, running tests to ensure the new changes do not introduce bugs, creating a patch in a version control system, submitting that patch via the project's preferred platform, waiting for someone in the project to review it, responding to any questions, and iteratively improving the patch as necessary so that it matches the project's contribution guidelines or reviewers' expectations.

For many potential contributors to documentation, these technical barriers pose a significant problem. We identified two kinds of technical barriers, which our interviewers either personally experienced or witnessed in cases of newcomers to a project:

*Using open source software platforms* Projects use many platforms, tools, and practices to manage their workflow of contributing code and documentation, each of which has its own learning curve. For example, newcomers must learn how a project uses a version control platform like GitHub and continuous integration platforms like Travis CI to submit, review, and incorporate changes. Furthermore, projects may also have differing community norms around contributing code (such as whether to rebase code before merging new contributions). As one interviewee noted, "there's not always consensus within the field about the right way to use those tools (Docathon organizer 2)."

*Using documentation-specific tools* There are also challenges in learning the tools that are specific to writing and building documentation. These tools require text to be formatted and structured in specific programmatic ways, which are often idiosyncratic to someone who isn't familiar with the tool. For example, putting the same information about a function in a python docstring can require writing different structured text, based on what tools are being used to automatically parse the text. The two code blocks below are docstrings that illustrate the difference between two popular formats: numpydoc and "Google-style":[2]

```python
def format_exception_numpy(etype, limit=None):
    """
    Format the exception with a traceback.

    Parameters
    ----------
    etype : str
        exception type
    limit : int or None
        maximum number of stack frames to show

    Returns
    -------
    out : list of strings
        list of strings
    """

def format_exception_google(etype, limit=None):
    """
    Format the exception with a traceback.

    Args:
        etype (str):  exception type

    Keyword Args:
        limit (int or None):  maximum number of stack frames to show

    Returns:
        out:  list of strings
    """
```

Some interviewees expressed concerns around technical barriers to newcomers, though noted that documentation is still often a good first-contribution for many people. Contributions to documentation generally will not "break" anything crucial in the package, are relatively easy to roll-back if an error is made, and provide an immediately-apparent contribution. One of our interviewees was a new contributor to open source software projects and the GitHub platform, and discussed their experience:

> And I learned a lot more about GitHub. I never had squashed or rebased before. Or I'd never really used branches correctly until that experience. So, I think it

---

[2] Adapted from https://bwanamarko.alwaysdata.net/napoleon/format_exception.html

> definitely made me better at using Git and ... a little more understanding of how open source is and like, the faces behind all the GitHub handles (Docathon participant 4)

### 2.3.3. Standards, quality assessment, and validation

One struggle many interviewees expressed around contributing documentation to open source software is the lack of standards and validation criteria for documentation. For example, in the previous subsection, we identified different documentation formats as a technical barrier. However, the many options for tooling introduces social challenges, as a there are widely differing opinions across communities on which standards should be used.

What constitutes good documentation is often contextual to various uses and goals, subjectively interpreted by different people, and left underspecified in community norms. This can especially be the case with tutorials, user guides, and other user/narrative documentation, rather than the typically well-structured and narrowly-scoped goals of API/reference documentation. Interestingly, some interviewees indicated that it was *more* difficult to contribute to user/narrative documentation (like tutorials or user guides) and much easier to contribute to examples or API documentation, which is generally highly structured. As one interviewee stated:

> Docstrings are supposed to be pretty terse and straightforward and those I'm not worried about doing on volunteer effort. Because again, basically, you take all the voice out and they say this is what it does, these are the parameters, this is what it returns. (Docathon participant 3)

In contrast to API documentation, user/narrative documentation can be complex and written with various narrative voices, points of view, or tenses. They have varying levels of structure, formality, and background knowledge assumed. They may also have inconsistencies in the author's tone, such as using humor or not. Consistent style and structure of documentation within a project was frequently identified as both an important property of good documentation as well as a major organizational challenge for open source software projects. Contributing user/narrative documentation can lead to long debates on details that have no one correct answer – often referred to as "bikeshedding" in OSS culture as inspired from Parkinson's law of triviality (Parkinson 1957).

Several interviewees discussed difficulties in getting pull requests around documen­men in generaltation accepted. One interviewee discussed frustration with getting their documentation contributions blocked because project developers objected to text they felt was "more like an opinion" (Docathon participant 4). Interviewees also mentioned "bikeshedding" around documentation. However, some stated that in some projects they felt it was easier to contribute documentation than code because it is "written rarely enough that people are very grateful that someone actually did that" (Open source contributor 10).

These tensions align with CSCW literature on conflict, particularly Hinds and Bailey's (2003) framework of task, process, and interpersonal conflict. Task conflict is when people disagree on what tasks ought to be done, process conflict is when people disagree about how the tasks ought to be accomplished, and interpersonal conflict centers around interpersonal relationships and interactional norms. These types are not mutually exclusive and one form of conflict can turn into another, but they help specify and distinguish different kinds of issues. We should also note that these issues are not unique to documentation, as they also frequently arise over code contribution.

2.4. Motivations for doing documentation work

Even if technical and social barriers were minimized in contributing documentation, an individual must still be motivated to do so. Another major theme in our interviews centered around incentives and credit (or the lack thereof) for doing documentation work. Our interviewees all believed that documentation was important and valuable for their projects, but there was a range of attitudes toward doing documentation work. In line with previous theoretical literature (Ryan and Deci 2000), we found it more useful to put interviewees' expressed motivations on a spectrum between fully intrinsic (where the task is seen as its own reward) to fully extrinsic (where the task is done for an external reward)—rather than see intrinsic/extrinsic as a binary. In Table 2 we outline Ryan and Deci's six kinds of motivations and give an example of each in the case of documentation work. We find it crucial to discuss motivations for doing documentation work in relation to motivations for other work in the OSS project, especially developing code. Most of our interviewees stated that documentation work in general was substantially less inherently enjoyable for them than developing code, which we discuss in the first subsection. In the second subsection, we discuss structural factors impacting motivation which differ between OSS projects, like project rules requiring documentation work or the level of credit/recognition for such work in the project.

*2.4.1. Do contributors enjoy doing documentation work in general?*
A large majority of our interviewees stated that documentation work is not as enjoyable for them, in the way that coding new features or fixing bugs is. This aligns with previous survey work finding that scientific open source software contributors enjoy tasks like writing code and fixing bugs far more than both writing and reviewing documentation (Holdgraf and Varoquaux 2017). Interviewees routinely used phrases like "eating your vegetables" or "bite the bullet", discussing how they felt it was important to write documentation for the good of the project, but that it was something they had to force themselves to do. Many of these interviewees also stated that this was a shared attitude among their peers, both in their own OSS projects and across OSS in general. "We all hate writing documentation" (Docathon participant 5), one interviewee stated matter-of-factly, adding that they were drawn to the idea of the Docathon because they felt it would facilitate some "team spirit" around a task that many people had neglected.

*Table 2.* Types of motivation with examples around cases of documentation work.

| Extrinsic or intrinsic | Kind | Example |
| --- | --- | --- |
| Fully extrinsic | Rewards | Getting paid to contribute documentation |
| Fully extrinsic | Regulation | Writing documentation because the project's rules require it |
| Slightly extrinsic | Introjection | Writing documentation in order to receive thanks, praise, recognition or to avoid guilt, shame, anxiety |
| Both extrinsic and intrinsic | Identification | Writing documentation because the community/project needs them to be successful, even though the work is not perceived enjoyable |
| Slightly intrinsic | Integration | Writing documentation because it is the right thing to do, getting a sense of satisfaction when completing documentation work |
| Fully intrinsic | Personal enjoyment | Writing documentation because the process is perceived as inherently enjoyable, fun, creative, etc. |

Several interviewees explicitly linked the issues around writing documentation to the contributor-driven nature of open source software development, stating that contributors contribute primarily to satisfy their own needs, making documentation a secondary goal. Several interviewees expressed what we call the "paradox of documentation:" those who know enough about the project to write documentation are the least in need of it:

> [writing documentation] is actually super hard. It's not super rewarding ... most people don't get the dopamine kick from writing documentation as implementing a new feature, right? Whether [you are adding] a new feature or you have a problem and you have fixed it, right? And the whole 'scratching my itch' aspect of open source typically means if you're working on something, you're working on something because it's bothering you. And you made it better and you're happy. Whereas with docs, the docs don't help you at all, because you know what they said because you wrote them. (Docathon participant 3)

However, two of our eleven interviewees (one Docathon organizer and one Docathon participant) did discuss the act of writing documentation as a creative process with high intrinsic enjoyment, similar to how they feel when writing code to develop new features or fix bugs. However, both of these interviewees also reflected that their attitudes were different than most in their communities. Both also stated they enjoyed and had extensive previous experience in other forms of writing, as well as having high competency in the English language.

Finally, a smaller number of interviewees expressed receiving strong levels of satisfaction from completing documentation tasks, such that they regularly performed such work — even though they did not generally inherently enjoy the task itself. One such interviewee made comparisons between documentation work and other forms of infrastructural and/or community support work that do not typically involve writing or fixing code directly, such as maintaining the build systems which automatically compile code to see if it runs on a variety of systems. They discussed their motivations to contribute to open source in general in terms of what would make the most impact, with work on documentation, build systems, releasing stable versions, and other meta-work having the biggest "return on investment" (Open source contributor 10) of their time.

*2.4.2. Structural factors relating to motivation*

Despite not inherently enjoying doing documentation work, most of our interviewees freely choose to do it without needing to be paid, forced, or shamed into doing it. We explore these different valences of motivation next, finding that motivation deeply intersects with various projects' specific organizational structures, cultural norms, as well as the peer production model of OSS projects. We find four structural factors that relate to motivation around documentation work: rules/policies requiring contributors to do documentation work, funding to pay contributors to do documentation work, contributors' feelings of responsibility toward users of a project, and the extent to which documentation work is valued and respected by other contributors as much as more 'technical' work like writing code.

*Rewards and rules* One of the most common extrinsic motivations to doing documentation work is being either directly paid to do such work or being required to do it in order to participate in the project. Some interviewees discussed projects that needed substantial overhaul in their documentation and hypothesized that it would only be done if someone was paid specifically to work on documentation. More foundations and grant agencies are awarding grants to specific OSS projects, particularly OSS libraries used for data analytics. Some grants awarded by funding agencies to support open source data analytics software projects specifically include documentation work as part of the tasks that will be done by those hired under the grant (Perez and Ganger 2015)—an emerging phenomenon that future research should investigate.

Interviewees also referenced projects which have documentation requirements as part of their rules around "pull requests," which is the process for submitting new changes to the codebase. Many open source projects in the data analytics ecosystem have increasingly standardized code review processes, especially for new features/functions. Much of the requirements are more code-focused, such as unit tests, conforming to code style guides like PEP8 in Python (Sharma et al.

2017), and passing a continuous integration check. Some of our interviewees discussed OSS libraries that have also added documentation requirements for new features, such that the project's code review rules do not permit a new feature to be added unless until it is documented (typically with API/reference documentation).

*Responsibility to users* We also heard many introjective motivations framed around responsibilities to users of the software library. Several of our interviewees discussed that an implicit responsibility of contributing to open source software project is in receiving requests from those who use the software. Even though there is no formal, contractual obligation to provide support, previous literature has discussed how open source software contributors feel obligations toward those who use software they have released (Lakhani and Wolf 2005; Kelty 2008). We similarly identified such issues in our study, particularly for more specialized libraries, which are common in the data analytics and scientific computing ecosystems. Some interviewees discussed projects where they were the only regular contributor and point of contact, regularly receiving direct requests from users. One sub-theme in interviewees' expressed motivations was around the perceived time it would save responding to questions from users on a more ad-hoc basis. Several interviewees referenced cases where either they or someone they personally knew (who did not enjoy writing documentation) ended up writing documentation because they were constantly receiving questions from users about how to use the software:

> The way the documentation got written there was the following:...they would send me an email [asking] ... how do you use it? So I would write a little explanation of how to do things. And after like, the fifth email, I was like "Well, maybe I should just make this a page." And once I made it a webpage ... well, maybe I should write a little bit of API documentation and a little bit of examples and so forth. And so, it was very kind of organic to where I got sick and tired of writing emails and I just put up a page. (Docathon participant 5)

*Recognition and credit* In larger and more popular OSS data analytics libraries, there are often dozens of regular contributors, and we found a common theme around community attitudes for documentation work. Many interviewees who regularly contribute documentation to such projects stated that they did not feel like they received same levels of positive community feedback for documentation work as they did for adding new features or fixing bugs. A common perception was that documentation work was perceived as being less valued, less important, and less "technical" than coding new features or fixing bugs. Participants discussed how documentation of a new or changed feature—which typically takes place after the coding work is complete—would often be de-prioritized, with developers moving on to other more "critical" tasks. This also varies from project to project, and some interviewees who contribute to multiple projects painted differing pictures of how much they believed these projects valued, respected, or even required documentation work. One interviewee discussed this perception, also raising issues with the gendered aspect of such work, stating that they did not want documentation work to be disproportionately performed by women—a theme long discussed by scholars of infrastructural and invisible work:

> One of the reasons open source documentation isn't great is it's definitely not viewed as as sexy as writing code. It's definitely viewed as less technical by some people [...]. And it's definitely viewed as less important by some people [...]. But that just kind of ties into the whole, you know, the trends everywhere of shunting women to work that is less valued by the community, type things. (Docathon participant 3)

In contrast, a smaller number of interviewees did feel like people in their projects were quite thankful when they wrote documentation. One explicitly referenced the perception that documentation is not valued as much as code, then took issue with it:

> I think there's this common perception about things that are not code ... let's say, documentation, is less valued than code. And especially people that write exclusively documentation are less valued than people that write also code or exclusively code. ... I would say that in general, this value system is not really true. I think on average, I've got way more positive responses on documentation contributions rather than code contributions, and I think that's true for other packages as well, because people do understand the value of documentation and especially because they don't like doing it, they're especially appreciative if you do it ... (Open source contributor 10)

These sentiments are likely to differ across projects (or as one interviewee noted, not so much between individual software projects, but between groups or ecosystems of projects that share the same developer community).

## 3. Discussion

The faces, roles, and practices of documentation in data analytics OSS libraries are inextricably linked to each other. The many different kinds, formats, and genres of documentation that proliferate in these projects are not accidental, but a result of the different roles that documentation plays for the many different people who use, produce, and maintain them. Similarly, the different kinds, formats, and genres of documentation we observed relate to our findings about how the work of producing and maintaining them is equally complex and multi-faceted. Even though documentation work is often cast as a less 'technical' activity, we find both traditionally 'social' and 'technical' skills and barriers in the practices of documentation. Motivations for doing documentation work are also linked to the different roles that documentation play for data analytics OSS libraries, as our many of interviewees' motivations were to do documentation work not for inherent enjoyment, but for the good of the project. But given that so many different kinds of documentation play so many different roles for so many different audiences, what it means to do documentation work for the good of the project also ranges widely.

3.1. What is 'good' documentation?

Throughout our interviews, documentation contributors to data analytics OSS libraries have not converged on a single, common definition of what "documentation" even is, much less what constitutes "good" documentation. We say this not as a criticism of these projects in the area of documentation, but rather to discuss how documentation is expected to serve many purposes, some of which may even be contradictory. Data analytics is changing as academic and industry researchers are moving away from commercial analytics software with graphical user interfaces and toward scripted programming languages extended by data analytics OSS libraries. Many of the growing pains of this transition are felt around documentation, which is the primary user interface for such software. Yet we also found that documentation for data analytics OSS libraries play a wide range of roles for a wide range of audiences and stakeholders, including but not limited to educating users about how to import and use this code for their own data analytics purposes. In addition, 'good' documentation is an advertisement for the features of the library, a signal of health for the community that develops the library, a living document capturing the institutional memory of features added or removed, a reference point for collaboration between developers of the library's code, a way to test and verify the integrity of new changes to the library's code, and as an onboarding mechanism for new contributors to the library's code.

There is no single format or genre of documentation that will serve all roles for all people, nor should there be. Communities should build common consensuses on defining what types of documentation they need as well as standards for what constitutes good and bad documentation. However, CSCW scholarship has shown the complexities with such standards (Bowker and Star 1999; Garfinkel1967). Standards and definitions will likely include more linguistic and narrative elements of style that are common across technical documentation and pedagogical material in general. Some qualities of good documentation will apply broadly across open source software libraries, though there will likely be qualities of good documentation that are particular to specific projects, software ecosystems, or programming languages. For example, based on our interviews, we suspect there are strong similarities, but also key differences in what constitutes good documentation in R versus Python or in a visualization library versus a machine learning library.

### 3.2. Demographics and inequalities in documentation work

While users and contributors largely agree upon the importance of documentation, the issue of who actually writes, reviews, and supports the work of documentation varies across communities. Based on our interviews, documentation is a kind of work that has implications for those concerned with systemic inequalities, particularly gender and those who do not speak English as a first language. CSCW researchers have studied similar phenomena across contexts, with theories like human infrastructure (Lee et al. 2006), invisible work (Suchman 1995; Star and Strauss 1999), and articulation work (Strauss 1988) that are quite applicable to our findings about documentation work in OSS data analytics libraries. CSCW has extensive methods and theories for studying how various forms of work are perceived as lower- and higher-status, as well as how different forms of work are more or less often performed by members of under-represented groups. Furthermore, with the largely volunteer-based nature of open source, the link between participation, recognition, and distribution of labor are crucial to issues of both diversity/inclusion and community sustainability – as previous researchers have investigated in Wikipedia (Menking and Erickson 2015).

While our study did not set out to specifically study issues of inequality and participation, several questions related to these areas arose during our research. Another key question for future research involves studying the extent to which documentation-related work is disproportionately performed by people from historically marginalized populations. Researchers and practitioners in open source software, scientific research, and peer production communities are increasingly concerns with inequalities, particularly around gender. Do people of different genders do documentation work at different rates? Is documentation work discussed with different kinds of gendered language compared to code work (e.g. as previous researchers have studied nursing versus surgical work in medicine)? Is it tackled more often by newcomers or veterans? By core or non-core members? Do these rates and proportions vary based on the type of documentation (e.g. API/reference docs, tutorials, galleries) or type of documentation work (e.g. writing vs. updating vs. reviewing)? Is documentation contributed by a small subset of the community, similar to findings that a small number of individuals generally contribute most of the content to open source software or Wikipedia (Muchnik et al. 2013)? How do these distributions change over time in response to various internal factors and external events, as well as across projects and languages?

### 3.3. Limitations and future work

Our study identified several common themes, which we discuss in this section. However, we must first note that our interviewees (and the projects we studied) are not a representative sample of open source software developers and do not intend for these claims to be interpreted as such. By design, our study sampled for diversity within the sub-population of contributors who already work on documentation within data analytics OSS projects. While we have not done strict cross-comparative work beyond this context, all the data analytics OSS libraries we observed are managed by an open peer-production community. In contrast, many OSS projects are developed by companies in more traditional mode of software engineering, then publicly released under open source licenses. As previous literature studying documentation in other contexts has shown, documentation work has different qualities and concerns, where dedicated professionals are often hired as technical writers to do this work.

As we have not studied documentation work in the much broader world of open source software in general, we are hesitant to make strong comparisons about whether documentation work in OSS libraries for data analytics are different than OSS libraries in general. Some themes we identified in our study have also been found in studies of how OSS developers in various projects interact with users in various social media channels (Storey et al. 2017). We do suspect that the heavy involvement of scientific researchers in data analytics libraries may impact the social organization of such communities, particularly as open source data analytics libraries are being championed by scientists concerned with issues of reproducibility. OSS libraries used for tasks like web development may have a very different set of issues and concerns, or substantial similarities. However, we do believe that OSS libraries raise a different set of issues around documentation when compared to standalone OSS applications, which have more traditional text or graphical user interfaces. OSS libraries extend the functionality of a programming language, and the the documentation is the primary user interface, rather than a GUI. We recommend that future research investigate the specific ways in which documentation work is similar and different across different kinds of OSS projects. Finally, we have largely investigated issues in English-language documentation, but there is substantial future work on multilingual and localized documentation. From our interviews and experiences, we suspect this work to be even more invisible than documentation work in English.

## 4. Conclusion

When we set out to study the Docathon, the questions we initially had were largely abstracted from the specifics of documentation work in data analytics OSS libraries. We had thought we would be focusing more on the "-thon" than the "doc-"—studying how people came together in a time-bounded, high-intensity event to do some kind of collaborative work in a computer-supported context. Yet as we dove deeper into studying the actual work around documentation in the context of data analytics OSS libraries, we found that documentation work raises a wide range of specific issues and concerns for those concerned with the social and technical infrastructures that support contemporary data analytics. Researchers and practitioners are increasingly focusing on the invisible, infrastructural work of developing and maintaining the code of open source software, both in general (Eghbal 2016) and specifically in academic research (Crouch et al. 2013). In this context, we find documentation work is invisible and infrastructural work twice-over: it is crucial to the continued operation and success of the invisible infrastructure of data analytics open source software libraries.

In this article, we have shown how software documentation in data analytics OSS libraries is a complex, multifaceted activity and artifact, irreducible to any single definition. Such a conclusion will likely not be surprising to scholars of documentation work in CSCW, but such an approach can be deeply generative for both scholars studying how data analytics is changing around computational methods, as well as for practitioners who are developing the social and technical infrastructures of contemporary data analytics. Documentation work is not just about describing what a particular data analytics library is, but also *why* it is— what is the project's goal, purpose, audience, scope, and mission? Similarly, in documentation work we see windows into the life of the community of people who develop and maintain it, as each of the types of documentation we observed has roles to play in creating and sustaining healthy and thriving communities. At the same time, there are a number of challenges associated with maintaining and growing quality documentation in this context. These include an imbalance in how credit and appreciation is given when it comes to documentation versus code, the necessity for different skill-sets when writing documentation, and various structural issues around the organization of peer production, open source software communities. These issues remain open areas for both CSCW research and interventions by practitioners in these communities, and we highly encourage future work in this area to involve similar collaborations between researchers and practitioners.

### Acknowledgments

We would like to thank our anonymous interviewees and Docathon participants for both their time in talking about documentation with us and in giving feedback on various drafts of this paper. We also thank Stacey Dorton for her extensive administrative assistance in our work, as well as the rest of the Docathon organizing team for their work this area: Stefan van der Walt, Dmitriy Morozov, Matthias Bussonnier, and Teon Brooks. Finally, we are grateful for the invaluable feedback of the anonymous peer reviewers in helping revise this paper.

### Funding Information

This work was supported by the Gordon and Betty Moore Foundation (Grant GBMF3834) and the Alfred P. Sloan Foundation (Grant 2013-10-27), as part of the Moore-Sloan Data Science Environments.

The Types, Roles, and Practices...

**Open Access**